\documentclass[11pt]{amsart}
\usepackage{amssymb}
\usepackage{graphicx}
\usepackage{epstopdf}
\newtheorem{theorem}{Theorem}

\newcommand{\CF}{\hbox{{$\mathcal F$}}}

\newcommand{\CL}{\hbox{{$\mathcal L$}}}

\newcommand{\cm}{\mathfrak{m}}  
\newcommand{\cg}{\mathfrak{g}}

\newcommand{\C}{\mathbb{C}}
\newcommand{\R}{\mathbb{R}}

\newcommand{\N}{\mathbb{N}}

\newcommand{\del}{\partial}

\newcommand{\h}{{\scriptstyle\frac{1}{2}}}

\newcommand{\extd}{{\rm d}}

\newcommand{\eps}{{\epsilon}}
\newcommand{\tens}{\mathop{\otimes}}
\newcommand{\la}{{\triangleright}}
\newcommand{\ra}{{\triangleleft}}

\newcommand{\Ad}{{\rm Ad}}

\newcommand{\id}{{\rm id}}
\newcommand{\<}{\langle}
\renewcommand{\>}{\rangle}

\def\rcross{{\triangleright\!\!\!<}}
\def\lcross{{>\!\!\!\triangleleft}}

\def\rlbicross{{\triangleright\!\!\!\blacktriangleleft}}
\def\lrbicross{{\blacktriangleright\!\!\!\triangleleft}}
\def\dcross{{\bowtie}}

\def\rbiprod{{\cdot\kern-.33em\triangleright\!\!\!<}}
\def\lbiprod{{>\!\!\!\triangleleft\kern-.33em\cdot}}

\newcommand{\eqn}[2]{\begin{equation}#2\label{#1}\end{equation}}

\begin{document}

\title[Noncommutative spacetimes]{Algebraic approach to quantum gravity II: noncommutative spacetime}
\author{Shahn Majid}\address{School of Mathematical Sciences\\
Queen Mary, University of London\\ 327 Mile End Rd,  London E1
4NS, UK\\ \& 
Perimeter Institute for Theoretical Physics\\
31 Caroline St N., Waterloo, ON N2L 2Y5, Canada}

\thanks{The work was completed on leave at the Mathematical Institute, Oxford, UK} 

\maketitle

\begin{abstract} We provide a self-contained introduction to the quantum group approach
to noncommutative geometry as the next-to-classical effective geometry that might be expected from any successful quantum gravity theory.  We focus particularly on a thorough account of  the bicrossproduct model noncommutative spacetimes of the form $[t,x_i]=\imath\lambda x_i$ and  the correct formulation of predictions for it including a variable speed of light. We also study global issues in the Poincar\'e group in the model with the 2D case as illustration. We show that any off-shell momentum can be boosted to infinite negative energy by a finite Lorentz transformaton.   \end{abstract}

\section{Introduction}

In this article we present noncommutative geometry (NCG ) not as a `theory of everything' but as a bridge between any future, perhaps combinatorial, theory of quantum gravity and the classical continuum geometry  that has to be obtained in some limit. We consider for the present that NCG is simply a more general notion of geometry that by its noncommutative nature should be the correct setting for the phenomenology and testing of  first next-to-classical quantum gravity corrections. Beyond that, the mathematical constraints of NCG may give us constraints on the structure of quantum gravity itself in so far as this has to emerge in a natural way from the true theory.

Also in this article we focus on the role of quantum groups or Hopf algebras\cite{Ma:book} as the most accessible tool of NCG, along the lines first introduced for Planck scale physics by the author in the 1980s \cite{Ma:pla,Ma:phy,Ma:mat,Ma:hop}. We provide a full introduction to our theory of `bicrossproduct quantum groups', which is one of the two main classes of quantum group to come out of physics (the other class, the $q$-deformation quantum groups, came out integrable systems rather than quantum gravity). The full machinery of noncommutative differential geometry such as gauge theory, bundles, quantum Riemannian manifolds, and spinors  (at least in principle) has also been developed over the last two decades; these topics are deferred to a third article in this series \cite{Ma:algIII}. This should allow the present article to be read without prior knowledge of either NCG or quantum groups. A first article in the series will be about the philosophical basis \cite{Ma:pri}. 

 As is well-known, quantum groups are a generalised notion of symmetry. There is a theorem that {\em all} bicrossproduct quantum groups indeed have associated to them noncommutative spaces on which they canonically act. Thus the  bicrossproduct  Poincar\'e quantum group denoted $U(so_{3,1})\rlbicross \C[\R^3\lcross\R]$ has associated to it  the proposal \cite{MaRue:bic} 
\eqn{R31}{ [x_0,x_i]=\imath\lambda x_i}
for a model of noncommutative 4D spacetime. Note that although (\ref{R31}) breaks usual Poincar\'e invariance, Special Relativity still holds as the quantum group `symmetry'. This is also the first  noncommutative spacetime model with a genuine physical prediction\cite{AmeMa:wav}, namely a  variable speed of light (VSL). The NASA GLAST satellite to be launched in 2007 may among other things be able to test this prediction through a statistical analysis of gamma-ray bursts even in the worst case that we might expect for the parameter $\lambda\sim 10^{-44}s$ ( the Planck timescale). Note that the model should not be confused with an earlier $\kappa$-Poincar\'e group model\cite{LNRT:def} where the quantum group had quite different generators (for example the Lorentz generators did not close among themselves so the physical interpretation was fundamentally different) and where prior to \cite{MaRue:bic} the spacetime on which it acts was assumed to be usual commutative Minkowski space (with nonsensical results). Similarly, the semidirect quantum group $U(su_2)\rbiprod_\lambda \C[SU_2]$ of Euclidean motions (a special case of a bicrossproduct called a Drinfeld double) acts
covariantly on 
\eqn{R3}{ [x_i,x_j]=2\imath\lambda\eps_{ij}{}^kx_k}
as noncommutative space or Euclideanised 3D spacetime \cite{BatMa:non}. Indeed this algebra arises in a certain limit as an effective description of Euclideanized 3D quantum gravity as proposed in \cite{BatMa:non} and recently proven in \cite{EL}.  It should not be confused with `fuzzy spheres' as we do not quotient to a matrix algebra or use any (in our opinion ad-hoc) matrix methods familiar in that context. One may also add a central $x_0$ to have a 4D spacetime\cite{Ma:tim}.
Notice that these and other noncommutative spacetimes in the paper are geometrically flat, i.e. they are relevant to a weak gravity regime of quantum gravity. Instead the effects they encode are of curvature in momentum space or `cogravity', a notion due to the author \cite{Ma:ista} as a potentially new and  independent physical effect. Due to running out of space, we will focus mainly on (\ref{R31}) and its illustrative 2D version, for which we provide a full global treatment.

Of course, the algebraic machinery that we shall describe includes many more models of potential physical interest. The bicrossproduct family nevertheless remain  the most interesting because they come from entirely classical (but non-linear) data. This means that although they are excellent examples of NCG their structure can be described ultimately by classical nonlinear differential equations and classical pictures. The classical data are a local factorisation of some Lie group $X\approx GM$ and equivalent to solving a pair of `matched pair' differential equations for an action of $G$ on $M$ and vice-versa. In \cite{Ma:pla} these were introduced  as toy models of Einstein's equations complete with `event-horizon-like' singularities; in the present application where the bicrossproduct is viewed as a Poincar\'e quantum group the latter appear as limiting asymptotes in momentum space, which has been called a `Planckian bound' on spatial momentum. This a generic feature of all bicrossproduct models based on noncompact groups. Moreover, the classical group $X=SO_{4,1}$  in the model (\ref{R31}) acts on the momentum group $M=\R^3\lcross\R$ and using this   action one can come up with an entirely classical picture equivalent to the model. The action of $G=SO_{3,1}$ is highly non-linear and given by certain vector fields in \cite{MaRue:bic}. We will demonstrate a new phenomenon for the model coming from this nonlinearity with explicit global formulae in the 2D case coming from $SO_{2,1}$.

Finally, a little knowledge can be a dangerous thing and certainly it is possible to claim any number of nonsensical `predictions' based on an abuse of the mathematics. If one is arguing as a phenomenologist then this does not matter; it does not matter where a formula comes from, one can just posit it and see if it fits the data. However,  for a  theoretical prediction one must have an actual theory. For this one has to address:

\begin{itemize} \item A somewhat complete mathematical framework within which to work  (in our case this will be NCG)
\item Is the proposal mathematically consistent?
\item What are {\em all} the physical consequences (is it physically consistent?)
\end{itemize}

Typically in NCG if one modifies one thing then many other things have to be modified for mathematical consistency (eg the Poincar\'e quantum group does not act consistently on ordinary spacetime). There will be many such issues adopting (\ref{R31}) and after that is the interpretation of the mathematics physically consistent? If we suppose that a symbol  $p^0$ in the mathematics is the energy then what else does this imply and is the whole interpretation consistent with other expectations? Or we can {\em suppose} that $p^\mu$ generators in the $\lambda$-Poincar\'e quantum group are the physically observed 4-momentum and from the deformed Casimir
\eqn{Cas}{||p||^2_\lambda=\vec p^2 e^{\lambda p^0}-{2\over\lambda^2}\cosh(\lambda p^0)-1)}
claim a VSL prediction but how to justify that?  Our approach is to look at noncommutative plane waves (or quantum group Fourier theory) to at least begin to turn such a formula  into a theoretical prediction \cite{AmeMa:wav}. The model (\ref{R31}) does then hold together  fairly well for scaler or $U(1)$ fields. Spinors in the model remain problemmatic and more theoretical development  would be needed before  predictions  involving neutrino oscillations or neutral kaon resonances etc. could have any meaning.
 
\section{Basic framework of NCG} \label{basic}

The framework that we use has the following elements. 

\begin{itemize} \item A spacetime coordinate algebra $A$, not necessarily commutative.

\item Differential calculus done algebraically as a linear map $\extd:A\to \Omega^1$ obeying some minimal axioms (here $\Omega^1$ is a bimodule of  `1-forms') . 

\item Symmetries done algebraically (e.g. as a quantum group)

\item An {\em Algebraic principle of equivalence:} All constructions are independent of any choice of generators of the algebras (the ability to change coordinates cf. passive diffeomorphism invariance in usual geometry). This does not mean that we might not prefer to work in some gauge such as in special relativity.

\item Insight into the new physics made possible by the particular framework.  In our case it is that nonocmmutative spacetime corresponds to a very natural idea: curved momentum space or {\em cogravity}.
\end{itemize}

Taking these in turn, we briefly define a differential calculus. This is common to all approaches to NCG except that in the quantum groups approach  one concentrates on $\Omega^1$ in the first instance. Requiring it to be an $A-A$ bimodule says that we can multiply `1-forms' by `functions' from the left or the right and the two associate:
\[ a((\extd b)c)=(a\extd b)c\quad\forall a,b,c\in A.\]
We also require that $\extd$ obeys the Leibniz rule
\[ \extd(ab)=a\extd b+ (\extd a)b\]
and that $\Omega^1={\rm span}\{a\extd b\}$ which is more of a definition than a requirement (if not we would just make $\Omega^1$ smaller). Finally there is an optional `connectedness' condition that
\[ \extd a=0\Rightarrow a\propto 1.\]
These axioms are all more or less obvious and represent the minimum that any form of geometry would require. They are actually {\em weaker} than classical differential geometry even when the algebra $A$ is commutative because we have not demanded anywhere that $[a,\extd b]=0$ for all $a,b$. Demanding that would imply that $\extd[a,b]=0$ for all $a,b,$ which would violate the connectedness condition for any reasonably noncommutative algebra.  Given $\Omega^1$ there are some different schemes to extend this to an entire exterior algebra $\Omega=\oplus_n\Omega^n$ with $\extd^2=0$, basically by some form of `skew-symmetrized' tensor products of 1-forms.

As soon as one has a calculus one can start to do physics, such as gauge theory at least at the level where a connection is a noncommutative (antihermitian) 1-form $\alpha$. Gauge transformations are invertible (unitary) elements $u$ of the noncommutative `coordinate algebra' and the connection and curvature transform as
\[ \alpha\to u^{-1}\alpha u+u^{-1}\extd u\]
\[F(\alpha)=\extd\alpha+\alpha\wedge\alpha \to u^{-1}F(\alpha)u.\]
Notice that the nonlinear term in $F$ does not automatically vanish since we did not assume that functions and 1-forms commute. Hence we call this $U(1)$-Yang-Mills theory to distinguish it from the Maxwell theory where $F=\extd \alpha$. The former detects noncommutative homotopy while the latter detects noncommutative de Rahm cohomology.

We do not actually need much from Hopf algebra theory other than the definitions and to be able to quote a couple of general results. A Hopf algebra or quantum group (we use the terms synonymously) means an algebra $H$ with unit which at the same time is a `coalgebra with counit' in a compatible way. By a coalgebra, say over $\C$, we mean
\[ \Delta:H\to H\tens H,\quad \eps:H\to \C\]
\[ (\id\tens\Delta)\Delta=(\Delta\tens\id)\Delta,\quad (\id\tens\eps)\Delta=(\eps\tens\id)\Delta=\id\]
(this is the same as the axioms of an algebra but with arrows reversed and $\Delta$ is called the  `coproduct', $\eps$ the `counit'. ) The compatibility with the algebra structure is that $\Delta,\eps$ should be algebra homomorphisms. In addition for a true quantum group there should exist a map $S:H\to H$ called the `antipode' such that
\[ \cdot(\id\tens S)\Delta=\cdot(S\tens\id)\Delta=1\eps.\]
If $H$ is a Hopf algebra then $H^*$ is at least an algebra with `convolution product' $(\phi\psi)(h)=(\phi\tens \psi)(\Delta h)$ for all $\phi,\psi\in H^*$. For suitable notions of dual it is again a quantum group, the dual one. If $H$ is a generalised symmetry algebra then $H^*$ is like the coordinate algebra on a generalised group. The basic `classical' example is when $H=U(\cg)$ the enveloping algebra of a Lie algebra. This is a Hopf algebra with
\[ \Delta x=x\tens 1+1\tens x,\quad\eps x=0,\quad Sx=-x,\quad \forall x\in\cg.\]
Its suitable dual is an algebra of coordinate functions $\C[G]$ on the associated Lie group. In the matrix Lie group case this is generated by matrix element  coordinates $\Lambda^\mu{}_\nu$ with  coproduct and counit
\[ \Delta \Lambda^\mu{}_\nu=\Lambda^\mu{}_\rho\tens \Lambda^\rho{}_\nu,\quad \eps \Lambda^\mu{}_\nu=\delta^\mu{}_\nu.\]
The antipode is given by matrix inversion. These two examples are all we need in most of the present article. 

For a Hopf algebra $H$ to act on an algebra $A$ we require that the product map $A\tens A\to A$ of the algebra is an intertwiner. The action of $H$ on $A$ extends to $A\tens A$ via the coproduct, so we require
\[ h\la (ab)= \cdot( (\Delta h)\la(a\tens b)),\quad h\la 1=\eps(h)1\]
where $h\la a$ denotes the action of $h$ on $a$ and $\la$ is similarly being used twice  on the right hand side of the first expression. For a calculus on $A$ to be covariant we require that $H$ acts on $\Omega^1$, that $\extd$ and the bimodule product maps are intertwiners. Part of the latter reads for example as
\[ h\la (a\extd b)=\cdot(\id\tens\extd)( (\Delta h)\la(a\tens b)).\]
Simply defining this as the action on $\Omega^1$ and knowing that it is well-defined implies the rest.
$H$ always acts on $H^*$ from both the left and the right by the coregular representation (e.g. the left action is  $h\la\phi=\phi((\ )h)$). In that case one can seek a calculus $\Omega^1$ on $H^*$  that is left and right covariant (bicovariant). This makes $H^*$ into the coordinate algebra of a  `quantum Lie group'. Note that one can work entirely with $H^*$ and never mention $H$ provided one uses the broadly equivalent notion of a `coaction' $\Delta_R:\Omega^1\to \Omega^1\tens H^*$ instead of an action $\la$ of $H$.

Similarly, an integral on an algebra $A$ just means a linear map $\int: A\to \C$. It is said to be $H$-covariant if
\[ \int (h\la a)=\eps(h)\int a,\quad \forall a\in A,\ h\in H\]
with respect to a covariant action $\la$ of $H$ on $A$. For a quantum group $A=H^*$ say (see above) if an $H$-covariant integral exists it is unique, cf. the Haar measure on a group. Again one can define it entirely with respect to $H^*$ if one uses the notion of a coaction.

The {\bf principle of algebraic equivalence} is the analogue of the statement in usual geometry that all constructions are covariant under coordinate change. This should not be confused with the physical equivalence principle, it is valid even in Newtonian mechanics and just says that we are free to change variables for  example from Cartesian to polar coordinates. This is what separates out the systematic  framework of NCG from `ad-hoc' constructions. This also makes clear why from our point of view  any argument for physical prediction based on casimirs in the Poincar\'e quantum group alone is completely empty. The reason is that most quantum groups including the bicrossproduct one for the spacetime (\ref{R31}) are {\em as algebras} isomorphic to the usual undeformed classical enveloping algebra. In other words there are new coordinates $P^\mu$ in which the quantum group is undeformed as an algebra and its Casimir is the usual $\vec P^2-(P^0)^2$. In this case the so-called prediction is like mistakenly working in polar coordinates while thinking they were Cartesian coordinates and being excited by the form of the Laplacian. In fact in the $P^\mu$ coordinates the coproduct of the quantum group also looks quite different but since the Casimir depends only on the algebra  it does not see this. Where the coproduct shows up is in tensor product actions of the quantum group (see above) and in truth the classical dispersion relation is not fully characterised by being a Casimir but by further properties in relation to this. Equivalently, how do we justify that $p^\mu$ in (\ref{Cas}) and not $P^\mu$ are the physical 4-momentum? The only way to know is to do experiments, and those experiments will likely involve objects such as plane waves that depend on the full quantum group structure not only the algebra. This means that early `predictions' based only on the algebra were wishful speculations and not theoretical predictions.  

On the topic of changing variables note that if $x_i$ are generators of $A$ then one might typically  have  $\extd x_i$ forming a basis over $A$ of $\Omega^1$. In this case the conjugate partial derivatives are defined by
\eqn{partial}{\extd a = \sum_i (\del^i a)\extd x_i.}
Notice that precisely when differentials do not commute with 1-forms, these $\del^i$ will not obey the usual Leibniz rule themselves. It is the coordinate-invariant object $\extd$ which obeys the Leibniz rule. Bases of $\Omega^1$ do not always exist and when they do they might not have the expected number, i.e. there might be additional auxiliary 1-forms beyond the classical basic 1-forms (see later).
Moreover, under a change of coordinates we leave $\extd$ unchanged and recompute the partial derivatives conjugate to the new basis. This is actually how it is done in classical differential geometry, only now we should do it in the noncommutive algebraic setting. The same remarks apply to the integral which will take a specific form when computed with one set of generators and another with a different set but with the same answer.

Finally, we promised one theorem and perhaps the most relevant is the quantum group Fourier transform \cite[paperback edn.]{Ma:book}. If $H,H^*$ are a dual pair of Hopf algebras (for some suitable dual) with dual bases $\{e_a\}$ and $\{f^a\}$ respectively, we define
\[ \CF:H\to H^*,\quad \CF(h)=\sum_a \int(e_a h)f^a,\quad \CF^{-1}(\phi)=S^{-1}e_a\int f^a\phi\]
where we assume the antipode $S$ is invertible (which is typical). This theory works nicely for finite-dimensional Hopf algebras but can also be applied at least formally to infnite-dimensional ones. Thus if $U(\cg)$ and $\C[G]$ mentioned above are {\em suitably completed} one has
at least formally
\[ \CF:\C[G]\to U(\cg),\quad \CF^{-1}:U(\cg)\to \C[G].\] 
The best approach here is actually to work with Hopf-von Neumann or $C^*$-algebra versions of these Hopf algebras.  For example $\C[G]$ might become continuous functions on $G$ with rapid decay at infinitiy in the noncompact case. The role of $U(\cg)$ might become the group $C^*$ -algebra which is a completion of the functions on $G$ with convolution product. However, we do not need to make this too precise at least for the bicrossproduct model. Formally we take a basis $\{\delta_u\}$ of $\delta$-functions on $G$ (more precisely one should smear or approximate these). For dual basis we take the group elements $u\in U(\cg)$ formally as exponential elements in the completed enveloping algebra. Then
\eqn{FTG}{ \CF(f)=\int_{G} \extd u f(u) u\approx \int_{U\subset\R^n} \kern -5pt \extd^nk\, J(k)f(k)e^{\imath k^ie_i}}
where $e_i$ are a basis of $\cg$ so that the $k^i$ are a local coordinate system for the group valid in some open domain $U$ and $J(k)$  the Jacobian for this change of variables.  There are subtleties particularly  in the compact case (e.g.  the case of $G=SU_2$ studied in detail in \cite{FreMa:non} as some kind of  `noncommutative sampling theory'). If  $G$ is a curved position space then the natural momenta $e_i$ are noncommuting covariant derivates and in the highly symmetric case of a nonAbelian group manifold they generate noncommutative momemtum  `operators' $U(\cg)$ instead of usual commutative coordinates. So actually physicists have been needing NCG -- in momentum space-- for about a century now, without knowing its framework. Indeed, Fourier transform is usually abandoned in any `functional' form on a nonAbelian group (instead one works  the whole category of modules, 3j and 6j-symbols etc.) but quantum group methods allow us for the first time to revert to Fourier transform as a functional transform, just with noncommutative functions $U(\cg)$. If this seems strange consider that the phase space of a particle on $G$ is $T^*G=\cg^*\times G$ and has quantum algebra of observables $U(\cg)\rcross \C[G]$ (in some form) -- this is called Mackey quantisation. Here $U(\cg)$ is contained in the algebra of obvervables as the quantisation $\C[\cg^*]$. This explains the top line in Figure~\ref{fig0}: gravity means noncommutative momentum space. Note that quantum mechanics itself is about cross relations between position and momentum as indicated for flat space in the bottom line of Figure~\ref{fig0}. We work in units where its associated variable $\hbar=1$.

  \begin{figure}\[\begin{array}{l|c|c}
&{\rm Position}& {\rm Momentum}\\
\hline
{\rm Gravity} & {\rm Curved}& {\rm Noncommutative} \\
& \sum_\mu x_\mu^2={1\over\gamma^2} & [p_i,p_j]=2\imath \gamma\eps_{ijk}p_k \\
\hline
{\rm Cogravity}& {\rm Noncommutative}& {\rm Curved}\\
& {}[x_i,x_j]=2\imath \lambda\eps_{ijk}x_k& \sum_\mu p_\mu^2={1\over\lambda^2}\\
\hline {\rm Quantum\ Mech.}&\multicolumn{2}{c}{
[x_i,p_j]=\imath\delta_{ij}}
\end{array}\]
\caption{\label{fig0}Noncommutative spacetime means  curvature in momentum space. The equations are for illustration.}
\end{figure}

On the other hand, now suppose that  $G$ is curved {\em momentum space} then the quantum group Fourier transform takes us equally well to a noncommutative enveloping algebra  $U(\cg)$ regarded as `coordinate functions' on some noncommutative position space. This is the exact form of (\ref{R31}) and (\ref{R3}) where $x_\mu$ or $x_i$ are the Lie algebra basis. So these noncommutative spacetimes are equivalent under quantum group Fourier transform to {\em classical} but curved momentum space.  This is the middle line in Figure~\ref{fig0}: noncommutativity in position space which should be interpreted as curvature in momentum space, i.e., the dual of  gravity or {\bf cogravity}. This is an independent physical effect and comes therefore with its own length scale which we denote $\lambda$.  These ideas were introduced in this precise form by the author in the mid 1990s on the basis of the quantum group Fourier transform\cite{Ma:ista}. Other works on the quantum group Fourier transform in its various forms include \cite{KemMa:alg, LyuMa:bra,LyuMa:qua}

\section{Bicrossproduct quantum groups and matched pairs}

We will give an explicit construction of the bicrossproduct quantum groups of interest, but let us start with a general theorem from the theory of Hopf algebras.  The starting point is a theory of factorisation of a group $X$ into subgroups $M,G$ such that $X=MG$. It means every element of $X$ can be uniquely expressed as a normal ordered product of elements in $M,G$. In this situation, define a left action $\la$ of $G$ on $M$ and a right action $\ra$ of $M$ on $G$ by the equation
\eqn{revprod}{us= (u\la s) (u\ra s),\quad \forall u\in G,\ s\in M.}
These actions obey
 \begin{eqnarray}\label{matchingcond}
u\ra e&=&u,\quad e\la s=s,\quad u\la e=e,\quad e\ra s=e \nonumber\\
(u\ra s)\ra t&=&u\ra (st),\quad u\la(v\la s)=(uv)\la s \nonumber\\
u\la (st)&=&(u\la s)((u\ra s)\la t)\nonumber \\
(uv)\ra s&=&(u\ra (v\la
s))(v\ra s)
\end{eqnarray}
for all $u,v\in G$, $s,t\in M$. Here $e$ denotes the relevant group unit element. A pair of groups equipped with such actions is said to be a `matched pair' $(M,G)$. One can then define   a `double cross
product group' $M\bowtie G$ with product
 \eqn{dcross}{(s,u).(t,v)=(s(u\la t), (u\ra t)v)}
and with $M,G$ as subgroups. Since it is built on the direct product space, the bigger group factorizes into these subgroups and in fact one recovers $X$ in this way.  These notions were known for finite groups since the 1910's but in a Lie group setting\cite{Ma:the,Ma:mat} one has the similar notion of a `local factoristion' $X\approx MG$ and a corresponding double cross sum $\cm\dcross \cg$ of Lie algebras. Then the differential version of the equations (\ref{matchingcond}) become a matter of a pair of coupled first order differential equations for  families of vector fields $\alpha_\xi$ on $M$ and $\beta_\phi$ on $G$  labelled by $\xi\in\cg$ and $\phi\in \cm$ respectively. We write these vector fields in terms of Lie-algebra valued functions $A_\xi\in C^{\infty}(M,\cm)$ and $B_\phi\in C^\infty(G,\cg)$ according to left and right translation from the tangent space at the identity:
\eqn{alphabeta}{ \alpha_\xi(s)=R_{s*}(A_\xi(s)),\quad \beta_\phi(u)=L_{u*}(B_\phi(u)).}
In these terms the matched pair equations become
\begin{eqnarray}\label{matchab}   A_\xi(st)=A_\xi(s)+\Ad_s(B_{\xi\ra s}(t)),\quad A_\xi(e)=0\nonumber
\\
B_\phi(uv)=\Ad_v^{-1}(A_{v\la \phi}(u))+B_\phi(v),\quad B_\phi(e)=0\end{eqnarray}
 along with auxiliary data a pair of linear actions $\la$ of $G$ in $\cm$ and $\ra$ of $M$ on $\cg$ exponentiating Lie algebra actions $\la,\ra$ of $\cg,\cm$ respectively. Finally, (\ref{matchab}) becomes a pair of differential equations if we let $u,t$ be infinitesimal i.e. elements $\eta\in\cg$, $\psi\in\cm$ say of the Lie algebra. Then
 \eqn{matchdif}{  \psi^R( A_\xi)(s)=\Ad_s( (\xi\ra s)\la \psi),\quad \eta^L(B_\phi)(v)=\Ad_{v^{-1}}( \eta\ra (v\la\phi))}
 where $\eta^L$ is the left derivative on the Lie group $G$ generated by  $\eta$ and $\psi^R$ the right derivative on $M$ generated by $\psi$. Note that this implies
 \eqn{initlara}{\psi^R( A_\xi)(e)=\xi\la \psi,\quad \eta^L(B_\phi)(e)=\eta\ra \phi}
 which shows how the auxiliary data are determined. These nonlinear equations were proposed in \cite{Ma:pla} as a toy model of Einstein's equations and solved for $\R\dcross\R$ where they were shown to have singularities and accumulation points not unlike a black-hole event horizon. Such accumulation points are a typical feature of (\ref{matchab}) when both groups are noncompact. We have flipped conventions relative to \cite{Ma:book} in order to have a left action of the Poincar\'e quantum group in our applications.

One has to solve these equations globally (taking account of any singularities)  in order to have honest Hopf-von Neumann or Hopf $C^*$-algebra quantum groups; there are some interesting open problems there. However, for simply a Hopf algebra at an algebraic level one needs only the initial data (\ref{initlara}) of the matched pair, namely the Lie algebra actions $\la,\ra$ corresponding to $\cm\bowtie\cg$. Clearly then $U(\cm\bowtie\cg)=U(\cm)\bowtie U(\cg)$ as a Hopf algebra double cross product or factorisation of Hopf algebras \cite{Ma:phy}. We content ourselves with one theorem from this theory: 

\begin{theorem}\label{bic} Let $(H_1,H_2)$ be a matched pair of quantum groups with $H_1\bowtie H_2$ the associated double cross product. Then (i) there is another quantum group denoted $H=H_2\rlbicross H_1^*$ called the `semidualisation' of the matched pair. (ii) This quantum group acts covariantly on $A=H_1$ from the left.  (iii) Its dual is the other semidualisation $H^*=H_2^*\lrbicross H_1$ and coacts covariantly on $H_1$ from the right.
\end{theorem}

Applying this theorem to $U(\cm)\dcross U(\cg)$ implies a bicrossproduct quantum group $U(\cg)\rlbicross U(\cm)$ acting covariantly on $A=U(\cm)$ from the left. Here it is assumed that $\C[M]$ is a suitable algebraic verison of the coordinate algebra of functions on $M$ dual to $U(\cm)$. The bicrossproduct quantum group is generated by $U(\cg)$ and the commutative algebra of functions on $M$, with cross relations and coproduct
\eqn{envrel}{ [f,\xi]=\alpha_\xi(f)}
\eqn{envcop}{ \Delta \xi=\xi \tens 1 + \Delta_L(\xi),\quad \Delta_L(\xi)\in \C[M]\tens \cg,\quad \Delta_L(\xi)(s)=\xi\ra s}
where $\Delta_L$ is the left coaction induced by the auxiliary linear action $\ra $ of $M$ on $\cg$. Meanwhile, the coproduct on $f\in \C[M]$ is that of $\C[M]$ which appears as a subHopf algebra. This is how we shall construct the bicrossproduct Poincar\'e quantum group enveloping algebra. Its canonical action on $U(\cm)$ from the theorem has $\xi\in \cg$ acting by the action $\la$ on $\cm$ and $f\in\C[M]$ acting by $(\id\tens f)\Delta$ using the coproduct of $U(\cm)$.

Equally, there is a natural dual  bicrossproduct as the Hopf algebra  $\C[G]\lrbicross U(\cm)$
coming from the same factorisation data. We denote by $a_\mu\in\cm$ the `nonAbelian translation' generators of $U(\cm)$ and by  $\Lambda^\mu{}_\nu$  any mutually commutative classical coordinates of the `Lorentz group' $G$ (as they will be in our application).  They obey
 \eqn{funrel}{{} [a_\rho,\Lambda^\mu{}_\nu]=\beta_{a_\rho}(\Lambda^\mu{}_\nu),\quad \Delta a_\mu=1\tens a_\mu+\Delta_R(a_\mu)}
 where $\beta$ is the other vector field in the matched pair and the coaction 
 \eqn{funcop}{  \Delta_R(a_\mu)=a_\nu\tens \Lambda^\nu{}_\mu\in \cm\tens \C[G],\quad \Delta_R(\phi)(u)=u\la \phi}
  is built similarly but now from $\la$ in the matched pair data. {\em By definition} the $\Lambda^\mu{}_\nu$ are the coordinate functions appearing in $\Delta_R$ on the $a_\mu$ basis. The construction, like (\ref{envrel})-(\ref{envcop}), is independent of any chosen generators but for Poincar\'e group coordinate functions one tends to use such notations.  If we denote by $x_\mu$ the `spacetime' generators of a second copy of $U(\cm)$ then the coaction of $\C[G]\lrbicross U(\cm)$ in Theorem~\ref{bic} is 
  \eqn{funcoa}{ \Delta_R ^{\rm Poinc}(x_\mu)=1\tens a_\mu +\Delta_R(x_\mu)=1\tens a_\mu+x_\nu\tens \Lambda^\nu{}_\mu.}

  In summary, the bicrossproduct theory constructs both the deformed Poincar\'e enveloping algebra and coordinate algebra at the same time and provides their canonical action and coaction respectively on another copy of $U(\cm)$ as noncommutative spacetime.

\subsection{Nonlinear factorisation in the 2D bicrossproduct model}

Such models provide noncommutative spacetimes and Poincar\'e quantum groups in any dimension $n$ based on a local factorisation of  $SO_{n,1}$ (de Sitter version) or $SO_{n-1,2}$ (ADS version). The 4D model is known\cite{MaRue:bic} but the 2D case has the same essential structure and we shall use this now to  explore global and nonlinear issues, with full derivations.

The first remark in the 2D case is that for a convenient description of the global picture we work  not with $SO_{2,1}$ exactly but its double cover $X=SL_2(\R)\to SO_{2,1}$. The map here at the Lie algebra level is
 \[  \tilde a_0={\lambda\over 2}\begin{pmatrix}1& 0\cr 0 & -1\end{pmatrix}\to \lambda \begin{pmatrix}0& 0& 1\cr 0&0 & 0\cr 1&0&0\end{pmatrix},\quad \tilde N={1\over 2} \begin{pmatrix}0& 1\cr 1 & 0\end{pmatrix}\to  {1\over \sqrt{2}}\begin{pmatrix}0& 0& 0\cr 0&0 & 1\cr 0&1&0\end{pmatrix}\] \[ \tilde a_1=\lambda \begin{pmatrix}0& 0\cr 1 & 0\end{pmatrix}\to{ \lambda\over \sqrt{2}}\begin{pmatrix}0& -1& 0\cr 1&0 & 1\cr 0&1&0\end{pmatrix}\]
 for $xt$, $yt$ boosts and $xy$-rotations  with  $++-$ signature generated by $-\imath\tilde a_0$, $\sqrt{2}\tilde N$, $\tilde M=-\imath\sqrt{2}(\lambda\tilde N-\tilde a_1)$ respectively.  The $\tilde a_i$ close to the Lie algebra $[\tilde a_1,\tilde a_0]=\lambda \tilde a_1$ so generate a 2-dimensional nonAbelian Lie group $M=\R\lcross\R$ along with $G=SO_{1,1}=\R$ generated by $\tilde N$. This gives a factorisation  $SL_2(\R)\approx (\R\lcross\R).SO_{1,1}$ as 
 \[  \begin{pmatrix}a& b\cr c & d\end{pmatrix}= \begin{pmatrix}a\mu& 0\cr {ac-bd\over a\mu} & {1\over a\mu}\end{pmatrix} \begin{pmatrix}{1\over\mu}& {b\over a\mu}\cr {b\over a\mu} & {1\over \mu}\end{pmatrix};\quad \mu=\sqrt{1-{b^2\over a^2}},\quad |b|<|a|.\]
 This is valid in the domain  shown which includes the identity in the group. It {\em cannot} be a completely global decomposition because topologically $SL_2(\R)$ and $PSL_2(\R)=SO_{2,1}$ have a compact direction and so cannot be described globally by 3 unbounded parameters (there is a compact $SO_2$ direction generated by $\tilde M$).  If one does not appreciate this and works with unbounded parameters one will at some stage encounter coordinate singularities, which is the origin of the Planckian bound  for this model as well as other new effects (see below). From an alternative constructive point of view, as we solve the  matched-pair equations for $(\R\lcross\R)\dcross\R$ we must encounter a singularity due to the nonlinearity.   Note that this nonAbelian factorisation and our construction of it cf.\cite{MaRue:bic} is not the KAN decomposition into three Abelian subgroups.

In the  factorisation we now change variables to
\[ a\mu=e^{{\lambda\over 2}p^0},\quad ac-bd=\lambda p^1e^{\lambda p^0},\quad \sinh({\theta\over 2})={b\over a\mu}\]
where we introduce $p^0,p^1$ as coordinates on the group  $M$ and $\theta$ as the coordinate of $SO_{1,1}$. Here $\lambda$ is a fixed but arbitrary normalisation constant and we have $\theta/2$ because we are working with the double cover of $SO_{2,1}$. According to the group law of matrix multiplication, the $p_i$ viewed abstractly as functions  enjoy the coproduct 
\[ \Delta \begin{pmatrix}e^{{\lambda\over 2}p^0}& 0\cr \lambda p^1e^{{\lambda\over 2} p^0}& e^{-{\lambda\over 2} p^0}\end{pmatrix}=\begin{pmatrix}e^{{\lambda\over 2}p^0}& 0\cr \lambda p^1e^{{\lambda\over 2} p^0}& e^{-{\lambda\over 2} p^0}\end{pmatrix}\tens \begin{pmatrix}e^{{\lambda\over 2}p^0}& 0\cr \lambda p^1e^{{\lambda\over 2} p^0}& e^{-{\lambda\over 2} p^0}\end{pmatrix}\]
where matrix multiplication is understood. Thus in summary we have 
\eqn{CM2}{[p^0,p^1]=0,\quad \Delta p^0=p^0\tens 1+1\tens p^0,\quad \Delta p^1=p^1\tens 1+ e^{-\lambda p^0}\tens p^1}
\eqn{CM2S}{ S(p^0,p^1)=(-p^0,-e^{\lambda p^0}p^1)}
as  the Hopf algebra $\C[\R\lcross\R]$ corresponding to our nonAbelian momentum group and its group inversion.

\begin{figure}\begin{picture}(240,240)(0,0)
 \put (0,0){\includegraphics{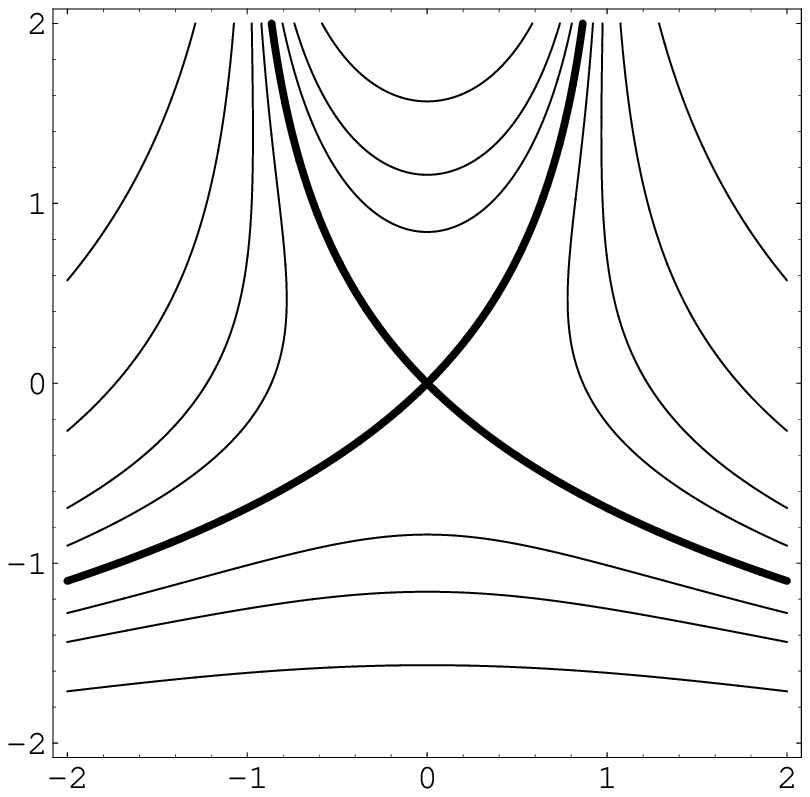}}
 \put(120,50){D}
  \put(120,170){A}
   \put(45,120){B}
    \put(195,120){C}
    \put(240,10){$\lambda p^1$}
    \put(-8,226){$\lambda p^0$}
  \end{picture}
 \caption{\label{fig1} Deformed orbits under the Lorentz group in the bicrossproduct model momentum group. Increasing $\theta$ moves anticlockwise along an orbit in regions A,D and clockwise in regions B,C.}  \end{figure}

We now take group elements in the wrong order and refactorise:
\[ \begin{pmatrix}\cosh({\theta\over 2})& \sinh({\theta\over 2}) \cr \sinh({\theta\over 2})& \cosh({\theta\over 2})\end{pmatrix}\begin{pmatrix}e^{{\lambda\over 2}p^0}& 0\cr \lambda p^1e^{{\lambda\over 2} p^0}& e^{-{\lambda\over 2} p^0}\end{pmatrix}=\begin{pmatrix}(C+S\lambda p^1)e^{{\lambda \over 2}p^0}& S e^{-{\lambda\over 2} p^0}\cr (S+C\lambda p^1)e^{{\lambda \over 2}p^0}& C e^{-{\lambda\over 2} p^0}\end{pmatrix} \]
\[=\begin{pmatrix}e^{{\lambda\over 2}p^0{}'}& 0\cr \lambda p^1{}' e^{{\lambda\over 2} p^0{}'}& e^{-{\lambda\over 2} p^0{}'}\end{pmatrix}\begin{pmatrix}\cosh({\theta'\over 2})& \sinh({\theta'\over 2}) \cr \sinh({\theta'\over 2})& \cosh({\theta'\over 2})\end{pmatrix} \]
where $S=\sinh(\theta/2)$, $C=\cosh(\theta/2)$, which gives according to (\ref{revprod}):
\eqn{p0'}{ p^0{}'= \theta\la p^0=p^0+{1\over\lambda}\ln\left((C+S\lambda p^1)^2-S^2e^{-2\lambda p^0})\right)}
\eqn{p1'}{ p^1{}'= \theta\la p^1\ra={(C+S\lambda p^1)(S+C\lambda p^1)-SC e^{-2\lambda p^0}\over \lambda\left( (C+S\lambda p^1)^2-S^2 e^{-2\lambda p^0}\right)}}
\eqn{theta'}{ \theta'=\theta\ra (p^0,p^1)=2{\rm arcsinh}\left({S e^{-{\lambda}p^0}\over\sqrt{(C+S\lambda p^1)^2-S^2 e^{-2\lambda p^0}}}\right)}
where we have written formulae in the domain where  $C+S\lambda p^1>0$. The refactorisation is possible (so the actions $\la,\ra$ are well-defined) only when 
\eqn{CSpatch}{\left (C+S(\lambda p^1-e^{-\lambda p^0})\right)\left (C+S(\lambda p^1+e^{-\lambda p^0})\right)>0.}
This can be analysed in terms of the regions in Figure~\ref{fig1} which shows orbits under $\la$ in $(p^0,p^1)$ space. One can check from the expressions above that these orbits are lines of constant values of 
\eqn{Cas2}{|| p||^2_\lambda = (p^1)^2e^{\lambda p^0}-{2\over\lambda^2}\left(\cosh(\lambda p^0)-1\right)={e^{\lambda p^0}\over\lambda^2}\left(\lambda^2(p^1)^2-(1-e^{-\lambda p^0})^2\right)}
which  deforms the Minkowski norm in momentum space. It is also invariant under inversion in the curved momentum group and hence under the antipode $S$. Note that this has nothing to do with the Poincar\'e algebra which we have not constructed yet; it is part of the nonlinear geometry arising from the factorisation.

\begin{theorem} (i) The actions $\la,\ra$ are defined for all $\theta$ if and only if $(p^0,p^1)$ lies in the upper mass shell (region A). \\ 
(ii) For any other $(p^0,p^1)$ there exists a  finite boost $\theta_c$ that sends  $p^0\to -\infty$, after which $\la$ breaks down. \\ (iii) For any $\theta$ there exists a critical curve not in region (A) such that approaching it sends $\theta\to \pm \infty$, after which $\ra$ breaks down. \end{theorem}

For the proof we use the shorthand  $q\equiv e^{-\lambda p^0}$. We  analyse the situation for the two cases $S>0$ and $S<0$; if $S=0$ then the condition (\ref{CSpatch}) always holds. Doing the first case, to lie in regions $A,C$ means  $\lambda p^1+1-q\ge 0$. Hence 
\[ C+S(\lambda p^1-q)=(C-S)+S(\lambda p^1+1-q)>0\]
which also implies that the other factor in (\ref{CSpatch}) is also positive, so the condition holds.  But conversely, strictly inside regions $B,D$ mean that $q-\lambda p^1>1$ and $C+S(\lambda p^1-q)=0$ has a solution $\theta_c>0$ according to ${\rm coth}({\theta_c\over 2})=q-\lambda p^1$.  We also note that our assumption $C+S\lambda p^1>0$ holds here and for all smaller $\theta$. As $\theta\to \theta_c$ from below, the denominator or  argument of log in the actions (\ref{p0'})-(\ref{p1'}) $\to 0$ and the transformed $p^0{}'\to -\infty$. If $S<0$ then $\lambda p^1+q-1\le 0$ in regions $A,B$ means that $C+S(\lambda p^1-q)>0$ and (\ref{CSpatch}) holds as before. Conversely, to be strictly inside regions $C,D$ means $\lambda p^1+q>1$ and hence $-\coth	{\theta_c\over 2}=\lambda p^1+q$ has a solution with $\theta_c<0$, where the denominators or argument of log again$\to 0$  from above as $\theta\to \theta_c$  from above. 

To give an example, consider a point  in  region D down from the 
origin, so $p^1=0$ and $p^0<0$. Then $e^{-\lambda p^0{}'}=e^{-\lambda p^0}/(1-\sinh^2({\theta\over 2})(e^{-2\lambda p^0}-1))$ blows up as $|\theta|\to | \theta_c|$ from below, where
\[ \theta_c=\pm  2{\rm arcsinh}\left({1\over \sqrt{ e^{-2\lambda p^0}-1}}\right)=\pm \ln{\rm coth}(-{\lambda\over 2}p^0).\]

\begin{figure}\begin{picture}(260,234)(0,0)
 \put (0,0){\includegraphics{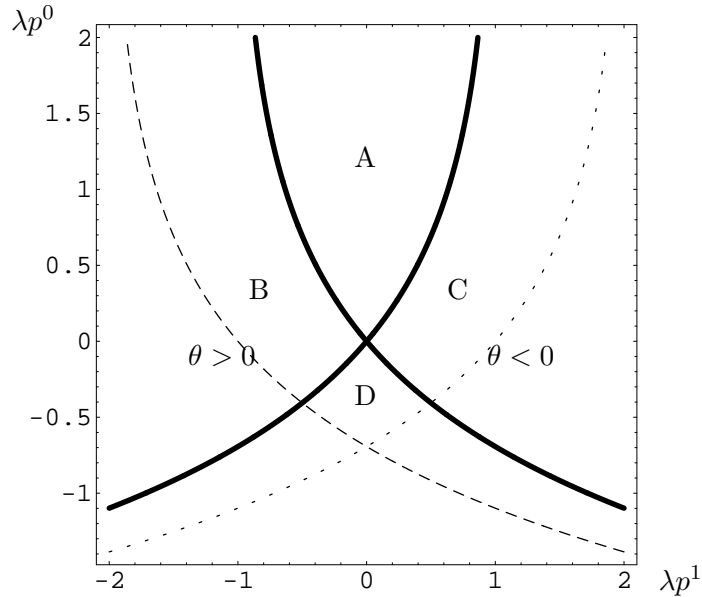}}
 \put(125,80){D}
  \put(125,170){A}
   \put(85,120){B}
    \put(160,120){C}
    \put(240,10){$\lambda p^1$}
    \put(-5,222){$\lambda p^0$}\put (175,95){$\theta<0$}\put(62,95){$\theta>0$}
  \end{picture}
 \caption{\label{fig2} Dashed (dotted) examples of critical curves for  given $\theta$. As $(p^0,p^1)$ approaches from above its action sends $\theta\to \pm\infty$.}  \end{figure} 

Pushing the arguments the other way, for any value  $\theta\ne 0$ we can clearly  find a critical curve of constant $q-\lambda p^1$ from the domains $B,D$ or of $q+\lambda p^1$ from $C,D$, according to the sign of $\theta$, such that  the same denominator factor, now in (\ref{theta'}), vanishes as we approach the critical curve from the origin. This is shown in Figure~\ref{fig2}.  The physical meaning of this will be given later as infinite uncertainty when this happens.

In summary, the nonlinearity behind the matched pair equations and the resulting action and back reaction between momentum and $SO_{1,1}$ has several consequences. We see in Figure~\ref{fig1} that the $p^0>0$ mass shells are now cups  with almost vertical walls, compressed into the vertical tube 
 \[ |p^1|< \lambda^{-1}.\]
In other words, the spatial momentum is bounded above by the Planck momentum scale (if $\lambda$ is the Planck time).  Indeed, this is immediate from (\ref{Cas2}). Such singularities expressed in accumulation regions  are a main discovery of the noncompact bicrossproduct theory visible already in the original examples\cite{Ma:pla}. They are a direct consequence of the nonlinearity but we also see their origin in the fact that the true group factorisation has a `curled up' compact direction. Moreover, this much-noted feature of the model is only a small part of the story. We see that fuller story is that any point outside this region is boosted to infinite negative $p^0$ by a finite boost with a similar story for $\theta$ and finite momentum as we saw in Figure~\ref{fig2}. Indeed the actions $\la,\ra$ breakdown at such points as the factorisation itself breaks down. Note also that the group inversion which is the natural reversal under CPT symmetry takes us from the `best' region $A$ to the `worst' region $D$, which is a {\bf  fundamental time-assymetry or non-reversebility} of the bicrossproduct model.
 
\subsection{Bicrossproduct $U_\lambda({\rm poinc_{1,1}})$ quantum group}

Now, consider $\theta$ infinitesimal, i.e. we differentiate all expressions (\ref{p0'})-(\ref{p1'}) by ${\del\over\del\theta}|_0$ which is all we need for the algebraic part of the bicrossproduct Hopf algebra (the full operator algebra structure needs the full global data). Thus denoting $N$ the Lie algebra generator conjugate to $\theta$, we have from the above the vector field and actions:
\[\alpha_{\imath N}={\del\over\del\theta}|_0=p^1{\del\over\del p^0}+{1\over 2}\left({1-e^{-2\lambda p^0}\over\lambda}-\lambda (p^1)^2\right){\del\over\del p^1}\]
\[ p^0\ra N=-\imath{\del p^0{}'\over\del\theta}|_{0}=-\imath p^1,\quad p^1\ra N=-\imath{\del p^1{}'\over\del\theta}|_{0}=-{\imath\over 2}({1-e^{-2\lambda p^0}\over \lambda}-{\lambda} (p^1)^2)\]
where the action $\ra$ flips to the other way because $\theta$ is really a coordinate function on $SO_{1,1}$ now being evaluated against $N$. A right-handed cross product by this action 
 gives the relations
\[ [p^0,N]=-\imath p^1,\quad [p^1,N]=-{\imath\over 2}\left({1-e^{-2\lambda p^0}\over\lambda}-\lambda (p^1)^2\right).\] 

Similarly differentiating the action (\ref{theta'})  {\em on} $\theta$ at $\theta=0$ gives the action of an element of $\R\lcross\R $ on $N$,  which we view equivalently as  coaction $\Delta_L$ of the coordinate algebra in algebraic terms, to find, 
\[  (p^0,p^1)\la N= e^{-\lambda p^0}N\  \Rightarrow\  \Delta_L(N)=e^{-\lambda p^0}\tens N\]  
which yields the coproduct and resulting antipode
\[  \Delta N=N\tens 1+ e^{-\lambda p^0}\tens N, \quad SN=-e^{\lambda p^0}N\]
to complete the structure of  $U_\lambda({\rm poinc}_{1,1})\equiv U(so_{1,1})\rlbicross \C[\R\lcross\R]$  along with (\ref{CM2}). 
Note that as $\lambda \to 0$ we obtain the 2D Poincar\'e algebra with the usual additive coproduct of $U({\rm poinc}_{1,1})$ as expected. Moreover, the deformed norm (\ref{Cas2}) is necessarily a constant of motion and hence killed by the vector $\tilde N$ (one may check this easily enough). Hence it is central (a Casimir) for the  deformed algebra.

In the 4D case the factorisation $SO_{4,1}\approx (\R^3\lcross\R).SO_{3,1}$ leading to Poincar\'e quantum group $U(so_{3,1})\rlbicross \C[\R^3\lcross\R]$ is too complicated to give explicitly but has similar global issues, likewise for $SO_{3,2}$. It was instead constructed in \cite{MaRue:bic} by  identifying the solution of the matched pair equations at the differentiated level as a result of finding the Hopf algebra itself  (we have seen that only the differentials of the actions $\la,\ra$ enter into the Hopf algebra itself) and integrating these. The Hopf algebra now has  commuting translation generators $p^\mu$, rotations $M_i$ and boosts $N_i$ with cf. \cite{MaRue:bic} but in opposite conventions for the coproduct:
\[   [p^\mu,p^\nu]=0,\quad [M_i,M_j]=\imath\eps_{ij}{}^kM_k,\quad 
[N_i,N_j]=-\imath\eps_{ij}{}^kM_k\] \[  [M_i,N_j]=\imath\eps_{ij}{}^kN_k,\quad [p^0,M_i]=0,\quad
[p^i,M_j]=\imath\eps^i{}_{j}{}_kp^k,\quad [p^0,N_i]=-\imath p_i,\]
as usual, and the modified relations and coproduct
\[ {}[p^i,N_j]=-{\imath\over 2}\delta^i_j \left({1-e^{-{2\lambda p^0}}\over
\lambda}+\lambda{\vec p}^2\right)+\imath\lambda p^ip_j,\]
\[  \Delta N_i= N_i\tens 1+e^{-{\lambda p^0}}\tens N_i+\lambda\eps_{ij}{}^kp^j\tens M_k,\]
\[ \Delta p^i= p^i\tens 1+ e^{-{\lambda p^0}}\tens p^i\]
 and the usual additive coproducts on  $p^0,M_i$.  The deformed Minkowski norm now has the same form as stated in (\ref{Cas}) with the same picture as in Figure~\ref{fig1} except that now the horizontal axis is any one of the $p_i$ (there is a suppressed rotational symmetry among them). As before, for the same fundamental  reasons of nonlinearity of the matched pair equations (\ref{matchab}), we have a Planckian bound $|\vec p|<\lambda^{-1}$ for particles on the $p^0>0$ mass-shell but we also have finite boosts sending off-shell or retarded momenta off to infinitely negative `energy'. 
 
 We have skipped over the 3D case, which is  of a similar form but without as many rotations and boosts. It was the first example in the general family to be found,  by the author in \cite{Ma:ista,Ma:the,Ma:hop,Ma:tim} as the bicrossproduct $U(so_3)\lrbicross\C[\R^2\lcross\R]$ (initially in a Hopf-von Neumann algebra setting),  from the 
 factorisation $SO_{3,1}= (\R^2\lcross\R).SO_3$. We have  similarly 
 \[   SO_{m,n}\approx (\R^{m+n-2}\lcross\R).SO_{m,n-1}\]
and it has been conjectured that the resulting bicrossproducts are all (nontrivially) isomorphic to certain contractions of the $q$-deformation quantum groups $U_q(so_{m,n})$.   In the 4D case the contraction of $U_q(so_{3,2})$ was found first \cite{LNRT:def} with the bicrossproduct construction found later in \cite{MaRue:bic}. Note that the physical interpretation of the generators coming from contractions is completely different from the bicrossproduct one.

\subsection{Bicrossproduct $\C_\lambda[{\rm Poinc}]$ quantum group}

We now apply the same matched pair factorisation data (\ref{p0'})--(\ref{theta'}) but now to construct the dual Hopf algebra. We start with $\C[SO_{1,1}]$ naturally described by  generators $s=\sinh({\theta})$ and $c=\cosh({\theta})$ with relations $c^2-s^2=1$  (which form the matrix $\Lambda^\mu{}_\nu$) and matrix coproduct
 \[  \Delta \begin{pmatrix}c&s\cr s&c\end{pmatrix}= \begin{pmatrix}c&s\cr s&c\end{pmatrix}\tens  \begin{pmatrix}c&s\cr s&c\end{pmatrix},\  \ S \begin{pmatrix}c&s\cr s&c\end{pmatrix}= \begin{pmatrix}c&-s\cr -s&c\end{pmatrix}.\]
To see how this arises in our theory, recall that we worked with $S=\sinh({\theta\over 2})$ and $C=\cosh({\theta\over 2})$ which (similarly) describe the double cover of $SO_{1,1}$ in coordinate form. We differentiate (\ref{theta'}) written in terms of $S$ by ${\del\over\del p^\mu}|_{p^\mu=0}$  to obtain the the vector fields $\beta$ and infinitessimal left action of the Lie algebra $[a_0,a_1]=\imath\lambda a_1$ on functions of $\theta$:
 \[ \beta_{\imath a_0}={\del\over\del p^0}|_0=-2\lambda CS {\del\over\del\theta},\quad \beta_{\imath a_1}={\del\over\del p^1}|_0=-2\lambda S^2{\del\over\del\theta}\]
  \[ a_0\la S=-\imath{\del\over\del p^0}|_0\sinh({\theta'\over 2})=\imath \lambda SC^2,\quad 
 a_1\la S=- \imath {\del\over\del p^1}|_0\sinh({\theta'\over 2})=\imath\lambda CS^2.\]
Note  also that  $\sinh(\theta)=2CS$ and $\cosh(\theta)=C^2+S^2$.  Hence from (\ref{funrel}) we find the relations 
\[ [a_0,\begin{pmatrix}c\cr s\end{pmatrix}]=\imath\lambda s \begin{pmatrix}s\cr c\end{pmatrix},\quad [a_1,\begin{pmatrix}c\cr s\end{pmatrix}]=\imath\lambda (c-1)\begin{pmatrix}s\cr c\end{pmatrix} \]
of the bicrossproduct $\C_\lambda[{\rm Poinc}_{1,1}]\equiv \C[SO_{1,1}]\lrbicross U(\R\lcross\R)$. Finally, differentiate (\ref{p0'})-(\ref{p1'}) to have the coaction $\Delta_R$ of $\C[SO_{1,1}]$ on the $a_\mu$:
 \[   {\del p^0{}'\over\del p^0}|_0=C^2+S^2={\del p^1{}'\over\del p^1}|_0,\quad {\del p^0{}'\over\del p^1}|_0=2CS={\del p^1{}'\over\del p^0}|_0\] 
 \[ \Rightarrow\quad \Delta_R(a_0,a_1)=(a_0,a_1)\tens  \begin{pmatrix}C&S\cr S&C\end{pmatrix}^2=(a_0,a_1)\tens  \begin{pmatrix}c&s\cr s&c\end{pmatrix}\]
 which along with the antipode completes the Hopf algebra structure constructed from (\ref{funrel})-(\ref{funcoa}). One can similarly describe the quantum group $\C_\lambda[{\rm Poinc}_{3,1}]=\C[SO_{3,1}]\lrbicross U(\R^3\lcross\R)$ in such a form, fitting in with a  classification of Poincar\'e coordinate quantum groups in a certain ansatz in \cite{PodWor}.

\section{Noncommutative spacetime, plane waves and calculus}

Until now we have given a quite technical construction of certain `Poincar\'e' Hopf algebras and spoken of `mass-shells' and  `energy' etc. but such appellations are meaningless until we consider the spacetime on which the algebra acts. Expressions such as (\ref{Cas2})  depend only on the algebra and can look however one wants depending on the arbitrary choice of generators named $p^\mu$. By contrast, the pair consisting of the quantum group {\em and} the spacetime on which it acts together have features independent of any choice of generators and this is where the actual physics lies as explained in Section~\ref{basic}. We turn to this now.

In the bicrossproduct models we know from Theorem~\ref{bic} that there is a canonical choice for this and it is noncommutative. Thus Poincar\'e quantum groups in the form (\ref{envrel})-(\ref{funcoa}) act (coact) on $U(\cm)$ and we recall that we denote the generators of this copy by $x_\mu$, which for the family above have the relations (\ref{R31}). We focus on the 4D case where $i=1,2,3$. The 3D case of these relations is the Lie algebra $\cm$ in  \cite{Ma:mat}.   

The first thing to do here is to explain the choice of momentum space coordinates in the previous section in terms of potentially physical quantities on this noncommutative spacetime, namely the noncommutative plane waves. The choice of momentum coordinates is arbitrary and as we change them the plane waves will look different. For our choice,
\[ \psi_{\vec p,p^0}=e^{\imath \vec p\cdot  x}e^{\imath
p^0 x_0},\quad 
\psi_{\vec p,p^0}\psi_{\vec p',p^0{}'}=\psi_{\vec p+e^{-\lambda p^0}\vec p',p^0+p^0{}'}\]
which shows the classical but nonAbelian group law of the Lie group $\R^3\lcross\R$ as read off from the product of plane waves.  It has exactly the same form as the coproduct (\ref{CM2}) before. Moreover, the quantum group Fourier transform reduces to the usual one but
 normal-ordered,
 \[ \CF(f)=\int_{\R^4}\extd^4 p\ f(p) e^{\imath \vec p\cdot \vec x}e^{\imath p^0 x_0}\]
and turns quantum differential operators on the noncommutative spacetime into multiplication operators. Put another way, the properly defined quantum differential operators will be diagonal on the noncommutative plane waves, as a general feature of all such models.
 
To complete the picture we need these quantum differentials, in order to describe the action of the $\lambda$-Poincar\'e generators on the noncommutative spacetime as differential operators. It is this action that physically specifies its role as `Poincar\'e' group to allow predictions. In the present model we have a natural differential calculus $\Omega^1$ with basis $\extd x_\mu$ and
\[ (\extd x_j)x_\mu=x_\mu\extd x_j,\quad (\extd
x_0)x_\mu-x_\mu \extd x_0=\imath\lambda \extd x_\mu\]
which leads to the partial derivatives
\eqn{del0}{
\del^i\psi=:{\del\over\del x_i}\psi(\vec x,x_0):=\imath p^i\la\psi}
\eqn{del1}{
\del^0\psi=:{\psi(\vec x,x_0+\imath\lambda)-\psi(\vec x,x_0)\over\imath \lambda}:={\imath\over\lambda}(1-e^{-\lambda p^0})\la\psi}
for normal ordered polynomial functions $\psi$ or in terms of the action of the momentum operators  $p^\mu$.  These $\del^\mu$ do respect our implicit $*$-structure (unitarity)  on the noncommutative spacetime but in a Hopf algebra sense which is not the usual sense since the action of the antipode $S$ is not just $-p^\mu$. This is fixed by adjusted derivatives $L^{-\h}\del^\mu$ where
\[ L\psi=:\psi(x,x_0+\imath\lambda):=e^{-\lambda p^0}\la \psi.\]
In this case the natural  4D Laplacian is $L^{-1}((\del^0)^2-\sum_i (\del^i)^2)$, which by (\ref{del0})-(\ref{del1}) acts on plane waves as (\ref{Cas}),  thereby  giving meaning to the latter as describing the physical mass-shell.

Finally, for the analysis of an experiment we assume the identification of noncommutive waves in the above normal ordered form with classical ones that a detector might register.  In that case  one may argue\cite{AmeMa:wav} that the speed for such waves  can be computed as $|{\del p^0\over\del p^i}|=e^{\lambda p^0}$ in units where 1 is the usual speed of
light.  So the prediction is that the speed of light depends on energy. What is remarkable is that
even if $\lambda\sim 10^{-44}s$ (the Planck time scale), this prediction could in principle be tested,
for example using $\gamma$-ray bursts.  These are known in some 
cases to travel cosmological distances before arriving here,
and have a spread of energies from 0.1-100 MeV. 
According to the above, the relative time delay $\Delta_T$ on
travelling distance $L$ for energies $p^0$, $p^0+\Delta_{p^0}$ is 
\[  \Delta_T\sim\lambda
\Delta_{p^0} {L\over c}\sim 10^{-44}{\rm s} \times 100 {\rm
MeV}\times  10^{10}{\rm y}\sim 1\ {\rm ms}\]
which is in principle observable by statistical analysis of a large number of
bursts correlated with distance (determined for example by using the Hubble telescope to
lock in on the host galaxy of each burst).  Although the above is only one of a class of predictions, it is striking that even Planck scale effects are now in principle within experimental reach.

\section{Physical interpretation}

We have given the bicrossproduct model  to the point of first predictions. However, there are still many issues for this and all other  models. The key problem is that in using NCG to model physics one still has to relate the mathematical objects to actual physics. That there is a fundamental issue here is evident in the following two questions:
 
 \begin{enumerate} \item How could we see a noncommutative plane wave? How would we   precisely measure any particular coordinates $p^\mu$ etc. labeling our plane waves. Without answering this, one has no prediction.
 
\item How would we physically detect the order of `addition' in the nonAbelian momentum group law? For example, if we smash together two waves of nonAbelian momentum $p,p'$, which way around to do we form the composite?

\end{enumerate}

\subsection{Preqantum states and quantum change of frames}

The correct way to address the first  issue according to current understanding is to treat the noncommutative algebra as an operator algebra, construct representations or `states'  of this `prequantum system' and consider that what would be observed macroscopically are expectation values 
$\< x_\mu \>$, $\< \psi_p(x)\>$ 
etc. in this state. Typically there exist `minimum uncertainty' coherent states  where the $x_\mu$ appear localised as much as possible around $ \< x_\mu  \>$ and the plane waves expectations in such coherent states have a specific signature that could be looked for, or conversely other states could be viewed as a superposition of these. For the model (\ref{R3}) see \cite{BatMa:non,Ma:tim}. In general  the  deeper theory of quantum gravity has  to provide these states and their behaviour {\em in addition} to the noncommutative spacetime and Poincar\'e algebra. Here $\lambda$ is treated as mathematically analogous to Planck's constant but is not Planck's constant (we work in units where $\hbar=1$), which is why we call this `prequantum' theory not quantum mechanics. It is something more fundamental. 

  Actually quantum gravity has to provide much more than this. It has to provide a representation of and hence expectation values for the entire coordinate algebra $\C_\lambda[{\rm Poinc}]$. Only given such a state would a quantum Poincar\'e transformation become an actual numerical transformation (as needed for example to pass to a rest frame) of the form
  \[ \<x_\mu \> \to \<x_\nu\> \<\Lambda^\nu{}_\mu\>+\<a_\mu\>+O(\lambda)\]
 where (say) the $a_\mu$ are the quantum group coordinates in the translation sector and $\Lambda^\mu{}_\nu$ are those in the Lorentz sector. In general one may not have such a decomposition, but even if one does, if one makes two such transformations, one will have in general that
 \eqn{aa}{  \<  a_\mu a_\nu\>= \<  a_\mu \>\< a_\nu\>+O(\lambda),\quad   \<  \Lambda^\mu{}_\nu a_\rho\>= \<  \Lambda^\mu{}_\nu\>\< a_\rho\>+O(\lambda)}
 \eqn{ll}{\<  \Lambda^\mu{}_\nu \Lambda^\alpha{}_\beta\>= \<  \Lambda^\mu{}_\nu\>\< \Lambda^\alpha{}_\beta\>+O(\lambda)}
reflecting that the quantum Poincar\'e coordinates do not commute in NCG; they are not given by actual numbers. NonAbelianness of the momentum group appears here in the first of  (\ref{aa}) which says that physical states provided by quantum gravity will not have classical numerical values for all the momentum coordinate operators $a_\mu$ simultaneously. This should not be confused with angular momentum (for example) where the enveloping algebra generators cannot be simultaneous diagonalised but where the coordinate algebra can be (actual classical values of angular momentum). Our situation is dual to that. We similarly cannot measure $\Lambda^\mu{}_\nu$ and $a_\rho$ simultaneously due to the second of (\ref{aa}) when the commutation relations are nontrivial. 

In the bicrossproduct model the $\Lambda^\mu{}_\nu$ mutually commute (the Lorentz coordinates are not deformed) so (\ref{ll}) does not need any $O(\lambda)$ corrections. States in this sector can be given by actual points in $SO_{3,1}$ or numerical angles. Meanwhile, the second of (\ref{aa}) has corrections due to (\ref{funrel})  given by the vector fields $\beta$ or in a global Hopf-von Neumann algebra setting by the global action $\la$ as in (\ref{theta'}), which we have seen blows up as in Figure~\ref{fig2}. This implies some form of `infinite nocommutativity' or `infinite uncertainty' for certain states. Thus,  while we have perfectly good Hopf algebras, they only see the differentiated data of the matched pair and miss the singular global picture. This enters when we try to represent them  as operator algebras in actual states. 

 In summary, a quantum Poincar\'e transformation makes sense algebraically but to realise it numerically one needs expectation values or representations of the generators of $\C_\lambda[{\rm Poinc}]$ (this is not be confused with representations of  $U_\lambda({\rm poinc})$ which have their usual meaning as  particle states). The lesson is that we need both in quantum gravity. 
  
 \subsection{$\bullet$-product, classicalisation and effective actions}
 
 An alternative approach to operator `prequantum' methods as above is to view the noncommutative spacetime algebra as a deformation on the same vector space as classically but with a new product $\bullet$. This comes with an identification $\phi$ of vector spaces, which we call the `classicalisation map', and which defines the modified product by
  \[ f\bullet g= \phi(\phi^{-1}(f)\phi^{-1}(g))\]
  for classical functions $f,g$.  We can add to this the {\bf working hypothesis} that noncommutative variables are to be observed by applying $\phi$ and observing the classical image. This brings with it a wealth of questions about why one should make such a postulate or what kind of supposition it makes about the experimental set up. In fact specifying $\phi$ is essentially equivalent to saying what one believes the noncommutative plane waves $\psi_p(x)$ look like, the implicit assumption being that these are to coincide under $\phi$  with their classical counterparts $e^{\imath p^\mu X_\mu}$ where we use $X^\mu$ for the classical spacetime coordinates. In that case
 \eqn{bulletwave}{e^{\imath p\cdot X}\bullet e^{\imath p'\cdot X}=\phi(\psi_p\psi_{p'})=\phi(\psi_{p p'})=e^{\imath (p p')\cdot X}}
 where $pp'$ denotes the (possibly nonAbelian) momentum group composition law in the chosen coordinate system. 

Thus for the bicrossproduct Minkowski spacetime the quantum plane waves above are equivalent to 
 \[ \phi(:f(x):)=f(X),\quad {\rm e.g.}\quad \phi(\psi_p(x))=e^{\imath p^\mu X^\mu}\]
 for any classical expression $f(X)$ and where $:\ :$ means putting all the $x_i$ to the left of all the $x_0$
 as explained in \cite{MaRue:bic}.  In experimental terms it means that experimental kit should (somehow) measure first $x_0$ and then $x_i$, the order mattering in view of the noncommutation relations. The bullet product implied here on classical functions is then
\eqn{bicbul}{f\bullet g=\cdot \left(e^{\imath\lambda  {\del\over\del X_0}\tens X_i{\del\over\del X_i}}(f\tens g)\right)=f(\vec X,X_0+\imath\lambda\deg(g))g(\vec X,X_0)}
 for classical functions $f,g$, where $\deg(g)$ is the total degree in the $X_i$ in the case where $g$ is homogeneous. Here one applies the operator shown and then multiplies the results using the classical product of functions on Minkowski space to give this result. This operator is a 2-cocycle in any Hopf algebra containing ${\del\over\del X_0}, X_i{\del\over\del X_i}$ which means it also fits into a `twist functor approach to quantisation'\cite{Ma:tan,MaOec:twi} leading to a different NCG  on the same algebra than the one from the bicrossproduct picture. We will not be able to cover the twist functor approach here due to lack of space but other twist functor models include the Moyal product or $\theta$-spacetime ({\em aka} the Heisenberg algebra) $[x_\mu,x_\nu]=\imath\theta_{\mu\nu}$.
 
Next, the classicalisation map allows one to write an NCG action like 
\eqn{CLqua}{ \CL=\int \extd f\wedge \star\extd f + m^2 f^2 +\mu  f^3 }
etc. where $f$ is an element of the quantum spacetime algebra and we assume we are given a covariant $\int$ and a Hodge $\star$-operator, in terms of ordinary fields $\phi=\phi(f)$ with action 
\eqn{CLclass}{\CL=\int_{\R^4} \extd^4X\, {\del\over\del X_\mu}\phi\bullet {\del\over\del X^\mu} \phi + m^2\phi\bullet\phi+ \mu \phi\bullet\phi\bullet\phi}
etc., using classical integration and calculus, but with the $\bullet$ product in place of the usual product of functions.   This assumes that $\int=\int\extd^4X \phi(\ )$ and that the quantum differentials become classical through $\phi$ as is the case for the simplest NCG models (including $\theta$-spacetimes and the 3D quantum double model (\ref{R3})). In the case of the bicrossproduct spacetime model the quantum integration is indeed defined by the normal ordering $\phi$ and we have seen (\ref{del0})-(\ref{del1}) that spatial quantum differentials indeed relate to the classical ones, but the $\del^0$ direction relates under $\phi$ to a finite difference in the imaginary time direction. Hence a noncommutative action will {\em not}  have a usual $\bullet$ form (\ref{CLclass}) but will involve finite differences for $\del^0$. One also has the problem that the quantum calculus and hence the NCG action is not necessarily $\lambda$-Poincar\'e covariant (even though the spacetime itself is), there is an anomaly for the Poincar\'e group at the differential level. One can replace the calculus by a 5D covariant one but then one has to interpret this extra direction. We expect it (see below) to relate to the renormalisation-group flow in the QFT on the spacetime. Again the physics of these issues remains fully to be explored at the time of writing.

\section{Other noncommutative spacetime models}

The 4D bicrossproduct model is the simplest  noncommutative spacetime model  that could be a deformation of our own world with its correct signature. There are less developed models and we  outline them here. 

We start with (\ref{R3}) for which $U_\lambda({\rm poinc}_{2,1})=U(so_{2,1})\rcross \C[SO_{2,1}]$ as a special case of a bicrossproduct where the back-reaction $\beta$ is trivial. Here $X=SO_{2,1}{}_{\rm Ad}\lcross SO_{2,1}$ and from the general theory we know that it acts on $U(so_{2,1})$ as a 3D noncommutative spacetime. Its Euclideanised version $U(su_2)$ is the algebra (\ref{R3})  proposed for 3D quantum gravity in \cite{BatMa:non}. For the plane waves, we use the canonical form  
\[ \psi_{\vec k}=e^{\imath  k\cdot x},\quad |\vec k|<{\pi\over\lambda}\]
in terms of the local `logarithmic' coordinates as in Section~\ref{basic}. The composition law for plane waves is the $SU_2$ product in these coordinates (given by the CBH formula) and we have a quantum Fourier transform (\ref{FTG}) with $e_i=x_i$ in the present application.  We also have  \cite{BatMa:non}:
\[\extd x_i={\lambda} \sigma_i,\quad x_i\Theta  -\Theta x_i= \imath{\lambda^2\over \mu}
\extd x_i,\]
\[ (\extd x_i)
x_j-x_j \extd x_i=\imath{\lambda} \eps_{ij}{}^k\extd
x_k+\imath{\mu}\delta_{ij}\Theta, \] where $\Theta$ is the $2\times 2$ identity matrix which,
together with the Pauli matrices $\sigma_i$ completes the basis of left-invariant 1-forms.  The 1-form  $\Theta$ provides a natural time direction, even though
there is no time coordinate, and the new parameter $\mu\ne 0$ appears as the freedom to change its normalisation. The partial derivatives $\del^i$ are defined by \[ \extd \psi(x)=(\del^i \psi)\extd
x_i+(\del^0 \psi)\Theta\]  and act diagonally on plane waves as
\[   \del^i=\imath {k^i\over\lambda |\vec k|} \sin({\lambda}|\vec k|),\quad \del^0=\imath{\mu\over\lambda}(\cos(\lambda |\vec k|)-1)=\imath{\mu\over 2}\vec\del^2+ O(\lambda^2).\]

Finally, there is a classicalisation map\cite{FreMa:non}
\[ \phi(\psi_{\vec k}(x))=e^{\imath p^\mu X_\mu},\quad p^0= \cos(\lambda |\vec k|),\quad p^i={\sin(\lambda |\vec k|)\over\lambda |\vec k|}k^i.\]
One can also label the noncommutative plane waves directly by $p^\mu$ as we did for the model (\ref{R31}). The map $\phi$  reproduces (\ref{R3}) by its $\bullet$ product and commutes with $\del_i$ (but not $\del^0$), which  means that actions such as (\ref{CLclass}) proposed in \cite{EL} as an effective theory for 3D quantum gravity essentially coincide with the NCG effective actions such as (\ref{CLqua}) as in \cite{BatMa:non}. Here $\int=\sum_{j\in \N}(j+1){\rm Tr}_j$ is the sum of traces in the spin $j/2$ representation. The noncommutative action has an extra term involving $\del^0$, which can be suppressed only by assuming that the 4D Hodge $*$-operator is degenerate.  Moreover, the map $\phi$ sees only the integer spin information in the model which is not the full NCG, see \cite{FreMa:non}.

Note that $\mu$ cannot be taken to be zero due to an anomaly for translation invariance of the DGA. This anomaly forces an extra dimension much as we saw for (\ref{R31}) before. The physical meaning of this extra direction $\del^0$ from the point of view of Euclidanized 3D quantum gravity is as a renormalisation group flow direction associated to blocking of the spins in the Ponzano-Regge model \cite{FreMa:non}. Alternatively, one can imagine this noncommuative spacetime arising in other nonrelativistic limits of a 4D theory, with the extra `time' direction  $x_0$ adjoined by \cite{Ma:tim}
\[ \Theta=\extd x_0,\quad [x_0,x_i]=0,\quad [x_0,\extd x_i]=\imath{{\lambda^2\over\mu}}\extd x_i,\quad [x_0,\Theta]=\imath{{\lambda^2\over\mu}}\Theta\]
and  new partial derivatives $\del^\mu$ on the extended algebra. Then the `stationary' condition in the new theory is $\extd\psi=O(\extd x_i)$ or  $\del^0\psi=0$, i.e. 
 \eqn{finsch}{  \psi(\vec x, x_0+\imath{\lambda^2\over\mu})=\left(\sqrt{1+{\lambda^2}\vec\del^2}\right)\psi(\vec x, x_0)}
which in the  $\lambda\to 0$ limit becomes the Schroedinger equation for a particle of mass $m=1/\mu$. Plane wave solutions exist in the form
\[ e^{\imath k^\mu x_\mu},\quad k^0=-{1\over m\lambda^2}\ln \cos(\lambda |\vec k|),\quad |\vec k|<{\pi\over 2\lambda}\]
showing the Planckian bound. 
  
Another major noncommutative spacetime, more or less fully explored by the author in the 1990's using braided methods is  $C_q[\R^{3,1}]$ or `$q$-Minkowski space'. It has a matrix of generators, relations, $*$-structure and braided coproduct
 \[\beta\alpha=q^2\alpha\beta,\quad
\gamma\alpha=q^{-2}\alpha\gamma,\quad \delta\alpha
=\alpha\delta,\]
\[\beta\gamma=\gamma\beta+(1-q^{-2})\alpha(\delta-\alpha),\]
\[\delta\beta=\beta\delta+(1-q^{-2})\alpha\beta,\ \gamma\delta=\delta\gamma+(1-q^{-2})\gamma\alpha,\]
 \[ \begin{pmatrix}\alpha&\beta\cr \gamma &\delta\end{pmatrix}^*=\begin{pmatrix}\alpha&\gamma \cr \beta &\delta\end{pmatrix},\quad \underline\Delta\begin{pmatrix}\alpha&\beta\cr \gamma &\delta\end{pmatrix}=\begin{pmatrix}\alpha&
\beta\cr \gamma &\delta\end{pmatrix}
\underline\tens \begin{pmatrix}\alpha&\beta\cr \gamma &\delta\end{pmatrix}\]
and is also denoted $B_q[M_2]$ as the algebra of braided $2\times 2$ Hermitian matrices \cite{Ma:exa}. If we quotient by the braided determinant relation $\alpha\delta-q^2\gamma\beta=1$ we have the unit hyperboloid in $\C_q[R^{3,1}]$ which is the braided group $B_q[SU_2]$  as obtained canonically from $\C_q[SU_2]$ by a process called `transmutation'.  Interestingly, the braided group is self-dual, $B_q[SU_2]\approx BU_q(su_2)=U_q(su_2)$ as an algebra, provided $q$ is generic; this is a purely quantum phenomenon. It means that $q$-Minkowski space has two limits, one is classical Minkowski space and the other after scaling and then taking the limit, is the enveloping algebra of $su_2\times u(1)$. There is also  an additive braided coproduct  $\underline\Delta \alpha=\alpha\tens 1+1\tens\alpha$, etc. which corresponds to the usual (flat) additive structure of $\R^{3,1}$. Finally, from braided group theory there is a  `bosonisation' construction $U_q({\rm poinc}_{3,1})=\widetilde{U_q(so_{3,1})}\rbiprod \C_q[\R^{3,1}]$ which acts covariantly on $\C_q[\R^{3,1}]$ as $q$-Poincar\'e quantum group with dilation\cite{Ma:mom}. Once again there is an anomaly which requires an extra generator, here a dilation indicated by $\widetilde{\ }$. It has been proposed that $q$-deformed models relate to quantum gravity with cosmological constant.


\baselineskip 14pt

 \begin{thebibliography}{99}
 
\bibitem[1]{AmeMa:wav}
G. Amelino-Camelia and S. Majid.
\newblock Waves on noncommutative space-time
and gamma ray bursts.
\newblock {\em Int. J. Mod. Phys. A} 15:4301--4324, 2000.

\bibitem[2]{BatMa:non} E. Batista and S. Majid.
\newblock Noncommutative geometry of angular momentum space $U(su_2)$.
\newblock {\em J. Math. Phys.} 44 (2003) 107-137.

\bibitem[3]{EL}
  L.~Freidel and E.~R.~Livine,
\newblock``Ponzano-Regge model revisited. III: Feynman diagrams and effective field theory,''
\newblock hep-th/0502106.

\bibitem[4]{FreMa:non}
S. Majid and L. Freidel.
\newblock Noncommutative harmonic analysis, sampling theory and the Duflo map in 2+1 quantum gravity, hep-th/0601004.

\bibitem[5]{KemMa:alg}
A. Kempf and S. Majid.
\newblock Algebraic q-Integration and Fourier Theory on Quantum and Braided Spaces.
\newblock {\em J. Math. Phys.} 35, 6802-6837, 1994.

\bibitem[6]{LyuMa:bra}
V. Lyubashenko and S. Majid.
\newblock Braided groups and quantum Fourier transform.
\newblock {\em J. Algebra} 166:506-528, 1994.

\bibitem[7]{LyuMa:qua}
V. Lyubashenko and S. Majid.
\newblock Fourier transform identities in quantum mechanics and the quantum line.
\newblock {\em Phys. Lett. B.} 284:66-70, 1992.

\bibitem[8]{LNRT:def}
J.~Lukierski, A.~Nowicki, H.~Ruegg, and V.N. Tolstoy.
\newblock {$q$}-{D}eformation of {P}oincar{\'e} algebra.
\newblock {\em Phys. Lett. B.} 268:331-338, 1991.

\bibitem[9]{Ma:ista}
S. Majid.
\newblock Duality principle and braided geometry, in {\em Springer Lect. Notes Phys}. 447:125--144, 1995.

\bibitem[10]{Ma:book} S. Majid.
\newblock  {\it Foundations of Quantum Group Theory},
C.U.P. 1995

\bibitem[11]{Ma:algIII}
S. Majid.
\newblock Algebraic approach to quantum gravity III: quantum Riemannian geometry.
\newblock To appear in B. Fauser and J. Tolksdorf, eds., Birkhauser.

 \bibitem[12]{Ma:the}
S.~Majid,
{\it Noncommutative-geometric Groups by a Bicrossproduct Construction},
(PhD thesis, Harvard mathematical physics, 1988).

 \bibitem[13]{Ma:pla}
S.~Majid.
\newblock {H}opf algebras for physics at the {P}lanck scale.
\newblock {\em J. Classical and Quantum Gravity}, 5:1587--1606, 1988.
 
\bibitem[14]{Ma:phy}
S. Majid.
\newblock Physics for algebraists: noncommutative and noncocommutative Hopf algebras by a bicrossproduct construction.
\newblock {\em J. Algebra} 130:17-64 (1990) 

\bibitem[15]{Ma:mat}
S.~Majid.
\newblock Matched pairs of {L}ie groups associated to solutions of the
  {Y}ang-{B}axter equations.
\newblock {\em Pac. J. Math.}, 141:311--332, 1990.

\bibitem[16]{Ma:hop}
S.~Majid.
\newblock {H}opf-von {N}eumann algebra bicrossproducts, {K}ac algebra
  bicrossproducts, and the classical {Y}ang-{B}axter equations.
\newblock {\em J. Funct. Analysis}, 95:291--319, 1991.

\bibitem[17]{Ma:pri}
S. Majid.
\newblock The Principle of Representation-theoretic Self-duality.
\newblock {\em Phys. Essays}, 4: 395-405, 1991.

\bibitem[18]{Ma:tan}
S. Majid.
\newblock Tannaka-Krein theorem for quasiHopf algebras and other results.
\newblock {\em Contemp. Math.} 134 (1992) 219-232.

\bibitem[19]{Ma:exa}
S. Majid.
\newblock Examples of braided groups and braided matrices.
\newblock {\em J. Math. Phys.} 32:3246-3253, 1991.

\bibitem[20]{Ma:mom}
S. Majid.
\newblock Braided momentum in the q-Poincar\'e group.
\newblock {\em J. Math. Phys.} 34:2045-2058, 1993.

 \bibitem[21]{Ma:tim}
S. Majid.
\newblock Noncommutative model with spontaneous time generation and Planckian bound.
 \newblock {\em J. Math. Phys.}  46 (2005) 103520, 18pp.

\bibitem[22]{MaOec:twi}
S. Majid and R. Oeckl.
\newblock Twisting of quantum differentials and the Planck scale Hopf algebra.
\newblock {\em Commun. Math. Phys.} 205: 617-655, 1999.

\bibitem[23]{MaRue:bic}
S.~Majid and H.~Ruegg.
\newblock Bicrossproduct structure of the {$\kappa$}-{P}oincar{\'e} group and
  non-commutative geometry.
\newblock {\em Phys. Lett. B}, 334:348--354, 1994.

\bibitem[24]{PodWor}
P. Podles and S. L. Woronowicz.
\newblock On the classification of quantum Poincar\'e groups.
\newblock {\em Comm. Math. Phys.} 178:61Ð82, 1996.

 \bibitem[25]{Wor:dif}
S.L. Woronowicz.
\newblock Differential calculus on compact matrix pseudpgroups (quantum groups). {\em Com. Math. Phys. } 122:125--170, 1989.

\end{thebibliography}
\end{document}